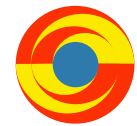

# Study on Linear Canonical Transformation in a Framework of a Phase Space Representation of Quantum Mechanics

Raoelina Andriambololona, Ravo Tokiniaina Ranaivoson, Rakotoson Hanitriarivo, Wilfrid Chrysante Solofoarisina

Theoretical Physics Dept., Institut National des Sciences et Techniques Nucléaires (INSTN-Madagascar), Antananarivo, Madagascar

**Email address:**
instn@moov.mg (Raoelina Andriambololona), raoelinasp@yahoo.fr (Raoelina Andriambololona), jacquelineraoelina@hotmail.com (Raoelina Andriambololona), tokhiniaina@gmail.com (R.T. Ranaivoson), infotsara@gmail.com (R. Hanitriarivo), s.wilfridc@moov.mg (W. C. Solofoarisina), wilfridc_solofoarisina@yahoo.fr (W. C. Solofoarisina)

**Abstract:** We present a study on linear canonical transformation in the framework of a phase space representation of quantum mechanics that we have introduced in our previous work [1]. We begin with a brief recall about the so called phase space representation. We give the definition of linear canonical transformation with the transformation law of coordinate and momentum operators. We establish successively the transformation laws of mean values, dispersions, basis state and wave functions. Then we introduce the concept of isodispersion linear canonical transformation.

## 1. Introduction

Because of the uncertainty relation [1],[2], studies about phase space and canonical transformations are among the most interesting problems in quantum theory. Many works have been done about these subjects since the beginning of the formulation of quantum mechanics and until nowadays [4],[5],[6],[7].

Let us consider one dimensional motion. Let $x$ be the position coordinate and $p$ the momentum and let $\boldsymbol{x}$ and $\boldsymbol{p}$ be respectively the coordinate and momentum operators [2]. We have the canonical commutation relation: $[\boldsymbol{x},\boldsymbol{p}]_{-} = i\hbar$. Operators are typed by a bold faced letter and the eigenvalues by a normal letter.

Let $|\psi\rangle$ be the quantum state vector and let $\psi(x) = \langle x|\psi\rangle$ and $\tilde{\psi}(p) = \langle p|\psi\rangle$ the corresponding wave functions respectively in coordinate and momentum representations. $\psi(x)$ and $\tilde{\psi}(p)$ are respectively the components of the vector $|\psi\rangle$ in the basis $\{|x\rangle\}$ and $\{|p\rangle\}$ of the state space. The mean values and statistical dispersions (or variances) of the coordinate and momentum are respectively

$$\langle x \rangle = \langle \psi|\boldsymbol{x}|\psi\rangle = \int x|\psi(x)|^2 \, dx \quad (1.1)$$

$$\langle p \rangle = \langle \psi|\boldsymbol{p}|\psi\rangle = \int p|\tilde{\psi}(p)|^2 \, dp \quad (1.2)$$

$$(\sigma_x)^2 = \langle \psi|(\boldsymbol{x}-\langle x\rangle)^2|\psi\rangle = \int (x-\langle x\rangle)^2|\psi(x)|^2 \, dx \quad (1.3)$$

$$(\sigma_p)^2 = \langle \psi|(\boldsymbol{p}-\langle p\rangle)^2|\psi\rangle = \int (p-\langle p\rangle)^2|\tilde{\psi}(p)|^2 \, dp \quad (1.4)$$

Using the properties of Fourier transform and Cauchy-Schwarz inequality, it may be shown the uncertainty relation [2].

$$\sigma_x \sigma_p \geq \frac{\hbar}{2} \quad (1.5)$$

Inequality (1.5) does not make easy the introduction of phase space in quantum mechanics which stipulates that exact values of coordinate and momentum cannot be determined simultaneously. However, many works have already been done concerning the formulation of quantum theory in phase space. Most of those works are based on the use of the Wigner quasi probability distribution introduced by Wigner in 1932 [4]. His pioneer work was completed by other authors such as Groenewold [5] and Moyal [6]. Their works had led to interesting results but the physical interpretation of Wigner distribution is not easy because it is not a positive definite distribution [11], [12].

In our work [1] we introduced another approach to tackle the problem of phase space in quantum mechanics with the introduction of a phase space representation. The goal of the present work is to perform a study of linear canonical transformation in the framework of this phase space representation. This latter is based on the introduction of a basis, in the state space, which is a set of phase space states denoted $|n,X,P,\Delta p\rangle$. A $|n,X,P,\Delta p\rangle$ is a state in which the mean values $X$, $P$ and statistical dispersions $(\Delta x_n)^2 = (2n+1)(\Delta x)^2$, $(\Delta p_n)^2 = (2n+1)(\Delta p)^2$ of coordinate and momentum are known, where $n$ is a nonnegative integer. $\Delta x$ and $\Delta p$ are linked by the relation [1]

$$(\Delta x)(\Delta p) = \frac{\hbar}{2} \quad (1.6)$$

The wave functions corresponding to a state $|n,X,P,\Delta p\rangle$ respectively in coordinate and momentum representations are the harmonic Gaussian functions $\varphi_n(x,X,P,\Delta p)$ and their Fourier transform $\tilde{\varphi}_n(p,X,P,\Delta p)$. These functions were introduced in our previous works [1], [3]. Their explicit expressions are

$$\langle x|n,X,P,\Delta p\rangle = \varphi_n(x,X,P,\Delta p) = \frac{H_n(\frac{x-X}{\sqrt{2}\Delta x})}{\sqrt{2^n n! \sqrt{2\pi}\Delta x}} e^{-(\frac{x-X}{2\Delta x})^2 + i\frac{Px}{\hbar}} \quad (1.7)$$

$$\langle p|n,X,P,\Delta p\rangle = \tilde{\varphi}_n(p,X,P,\Delta p) = \frac{(-i)^n H_n(\frac{p-P}{\sqrt{2}\Delta p})}{\sqrt{2^n n! \sqrt{2\pi}\Delta p}} e^{-(\frac{p-P}{2\Delta p})^2 - iX\frac{(p-P)}{\hbar}} \quad (1.8)$$

And we have the relations

$$X = \langle n,X,P,\Delta p|\mathbf{x}|n,X,P,\Delta p\rangle = \int x|\varphi_n(x,X,P,\Delta p)|^2 \, dx \quad (1.9)$$

$$P = \langle n,X,P,\Delta p|\mathbf{p}|n,X,P,\Delta p\rangle = \int p|\tilde{\varphi}_n(p,X,P,\Delta p)|^2 \, dp \quad (1.10)$$

$$(\Delta x_n)^2 = \langle n,X,P,\Delta p|(\mathbf{x}-X)^2|n,X,P,\Delta p\rangle = \int (x-X)^2|\varphi_n(x,X,P,\Delta p)|^2 \, dx \quad (1.11)$$

$$(\Delta p_n)^2 = \langle n,X,P,\Delta p|(\mathbf{p}-P)^2|n,X,P,\Delta p\rangle = \int (p-P)^2|\tilde{\varphi}_n(p,X,P,\Delta p)|^2 \, dp \quad (1.12)$$

The phase space state $|n,X,P,\Delta p\rangle$ are eigenstates of operators called dispersion operators namely: coordinate dispersion operator $\Sigma_x$ and momentum dispersion operator $\Sigma_p$.

$$\begin{cases} \boldsymbol{\Sigma}_x = \frac{1}{2}[(\mathbf{x}-X)^2 + \frac{(\Delta x)^2}{(\Delta p)^2}(\mathbf{p}-P)^2] \\ \boldsymbol{\Sigma}_p = \frac{1}{2}[(\mathbf{p}-P)^2 + \frac{(\Delta p)^2}{(\Delta x)^2}(\mathbf{x}-X)^2] \end{cases} \quad (1.13)$$

The eigenvalues of the operators $\boldsymbol{\Sigma}_x$ and $\boldsymbol{\Sigma}_p$ are equal, respectively, to $(\Delta x_n)^2$ and $(\Delta p_n)^2$.

$$\begin{cases} \boldsymbol{\Sigma}_x|n,X,P,\Delta p\rangle = (\Delta x_n)^2|n,X,P,\Delta p\rangle \\ \boldsymbol{\Sigma}_p|n,X,P,\Delta p\rangle = (\Delta p_n)^2|n,X,P,\Delta p\rangle \end{cases} \quad (1.14)$$

with $(\Delta x_n)^2 = (2n+1)(\Delta x)^2$, $(\Delta p_n)^2 = (2n+1)(\Delta p)^2$. Relations (1.7) and (1.8) permit to perform the passage between coordinate and momentum representations, basis $\{|x\rangle\}$ and $\{|p\rangle\}$, and phase space representation basis $\{|n,X,P,\Delta p\rangle\}$). For any state $|\psi\rangle$, and by setting $\psi(x) = \langle x|\psi\rangle, \tilde{\psi}(p) = \langle p|\psi\rangle$ and $\Psi^n(X,P,\Delta p) = \langle n,X,P,\Delta p|\psi\rangle$ respectively the wave functions in coordinate, momentum and phase space representations, we have shown the relations [1]

$$\Psi^n(X,P,\Delta p) = \langle n,X,P,\Delta p|\psi\rangle = \int \langle n,X,P,\Delta p|x\rangle\langle x|\psi\rangle dx = \int \varphi_n^*(x,X,P,\Delta p)\psi(x)dx$$

$$= \int \langle n,X,P,\Delta p|p\rangle\langle p|\psi\rangle dp = \int \tilde{\varphi}_n^*(p,X,P,\Delta p)\tilde{\psi}(p)\, dp \quad (1.15)$$

$$\psi(x) = \langle x|\psi\rangle = \int \langle x|n,X,P,\Delta p\rangle\langle n,X,P,\Delta p|\psi\rangle \frac{dXdP}{2\pi\hbar} = \int \Psi^n(X,P,\Delta p)\, \varphi_n(x,X,P,\Delta p)\frac{dXdP}{2\pi\hbar}$$

$$= \sum_n \langle x|n,X,P,\Delta p\rangle\langle n,X,P,\Delta p|\psi\rangle = \sum_n \Psi^n(X,P,\Delta p)\, \varphi_n(x,X,P,\Delta p) \quad (1.16)$$

$$\tilde{\psi}(p) = \langle p|\psi\rangle = \int \langle p|n,X,P,\Delta p\rangle\langle n,X,P,\Delta p|\psi\rangle \frac{dXdP}{2\pi\hbar} = \int \Psi^n(X,P,\Delta p)\, \tilde{\varphi}_n(p,X,P,\Delta p)\frac{dXdP}{2\pi\hbar}$$

$$= \sum_n \langle p|n,X,P,\Delta p\rangle\langle n,X,P,\Delta p|\psi\rangle = \sum_n \Psi^n(X,P,\Delta p)\, \tilde{\varphi}_n(p,X,P,\Delta p) \quad (1.17)$$

In the present work, we look to study linear canonical transformation in the framework of the phase space representation which has been introduced in our paper [1] and recalled briefly here-above. The definition of the linear canonical transformations is given in section 2 with the transformation laws for coordinate and momentum operators. In the section 3, we establish the transformation laws for mean values and dispersions. Section 4 is focused on the study of the transformation equations for the eigenstates, of the coordinate and momentum operators, and wave functions. In section 5, the concept of isodispersion linear canonical transformations is introduced. We conclude in the section 6.

## 2. Linear Canonical Transformation

In the framework of quantum mechanics, a linear canonical transformation can be defined as a linear transformation mixing coordinate and momentum operators $x$ and $p$ and leaving invariant the canonical commutation relation. For convenience, we use the phase space representation and we write the transformation in the form

$$\begin{cases} y = ax + b\frac{\Delta x}{\Delta p}p \\ k = c\frac{\Delta p}{\Delta x}x + dp \end{cases} \quad (2.1)$$

$$[y, k]_- = [x, p]_- = i\hbar \quad (2.2)$$

Combining (2.1) and (2.2), we obtain the condition which links the parameters, $a, b, c$ and $d$, of the linear canonical transformation

$$ad - bc = 1 \quad (2.3)$$

Because of (2.3), there are only three independent parameters.

## 3. Transformation Laws of Mean Values and Dispersion

For a state $|\psi\rangle$, the corresponding mean values and dispersions of the coordinate and the momentum are given by the relations

$$\langle x \rangle = \langle \psi | x | \psi \rangle \quad (3.1)$$

$$\langle p \rangle = \langle \psi | p | \psi \rangle \quad (3.2)$$

$$(\sigma_x)^2 = \langle \psi | (x - \langle x \rangle)^2 | \psi \rangle \quad (3.3)$$

$$(\sigma_p)^2 = \langle \psi | (p - \langle p \rangle)^2 | \psi \rangle \quad (3.4)$$

For future use, we define also the coordinate-momentum and momentum-coordinate codispersions corresponding to the state $|\psi\rangle$, respectively as follows

$$\sigma_{xp} = \langle \psi | (x - \langle x \rangle)(p - \langle p \rangle) | \psi \rangle \quad (3.5a)$$

$$\sigma_{px} = \langle \psi | (p - \langle p \rangle)(x - \langle x \rangle) | \psi \rangle \quad (3.5b)$$

Since the operators $x$ and $p$ do not commute, we have $\sigma_{xp} \neq \sigma_{px}$, but we have the relation

$$\sigma_{xp} - \sigma_{px} = \langle \psi | (x - \langle x \rangle)(p - \langle p \rangle) - (p - \langle p \rangle)(x - \langle x \rangle) | \psi \rangle = \langle \psi | i\hbar | \psi \rangle = i\hbar$$

If we consider the linear canonical transformation (2.1) we can deduce for the new variables

$$\langle y \rangle = \langle \psi | y | \psi \rangle = \langle \psi | ax + b\frac{\Delta x}{\Delta p}p | \psi \rangle = a\langle x \rangle + b\frac{\Delta x}{\Delta p}\langle p \rangle \quad (3.6)$$

$$\langle k \rangle = \langle \psi | k | \psi \rangle = \langle \psi | c\frac{\Delta p}{\Delta x}x + dp | \psi \rangle = c\frac{\Delta p}{\Delta x}\langle x \rangle + d\langle p \rangle \quad (3.7)$$

$$(\sigma_y)^2 = \langle \psi | (y - \langle y \rangle)^2 | \psi \rangle = \langle \psi | [(ax + b\frac{\Delta x}{\Delta p}p) - (a\langle x \rangle + b\frac{\Delta x}{\Delta p}\langle p \rangle)]^2 | \psi \rangle$$

$$= a^2(\sigma_x)^2 + ab\frac{\Delta x}{\Delta p}(\sigma_{xp} + \sigma_{px}) + b^2(\frac{\Delta x}{\Delta p})^2(\sigma_p)^2 \quad (3.8)$$

$$(\sigma_k)^2 = \langle \psi | (k - \langle k \rangle)^2 | \psi \rangle = \langle \psi | [(c\frac{\Delta p}{\Delta x}x + dp) - (c\frac{\Delta p}{\Delta x}\langle x \rangle + d\langle p \rangle)]^2 | \psi \rangle = c^2(\frac{\Delta p}{\Delta x})^2(\sigma_x)^2 + cd(\sigma_{xp} + \sigma_{px}) + d^2(\sigma_p)^2 \quad (3.9)$$

$$\sigma_{yk} = \langle \psi | (y - \langle y \rangle)(k - \langle k \rangle) | \psi \rangle = \langle \psi | \left[(ax + b\frac{\Delta x}{\Delta p}p) - (a\langle x \rangle + b\frac{\Delta x}{\Delta p}\langle p \rangle)\right]\left[(c\frac{\Delta p}{\Delta x}x + dp) - (c\frac{\Delta p}{\Delta x}\langle x \rangle + d\langle p \rangle)\right] | \psi \rangle$$

$$= ac\frac{\Delta p}{\Delta x}(\sigma_x)^2 + ad\sigma_{xp} + bc\sigma_{px} + bd\frac{\Delta x}{\Delta p}(\sigma_p)^2 \quad (3.10)$$

$$\sigma_{ky} = \langle \psi | (k - \langle k \rangle)(y - \langle y \rangle) | \psi \rangle = \langle \psi | \left[(c\frac{\Delta p}{\Delta x}x + dp) - (c\frac{\Delta p}{\Delta x}\langle x \rangle + d\langle p \rangle)\right]\left[(ax + b\frac{\Delta x}{\Delta p}p) - (a\langle x \rangle + b\frac{\Delta x}{\Delta p}\langle p \rangle)\right] | \psi \rangle$$

$$= ac\frac{\Delta p}{\Delta x}(\sigma_x)^2 + bc\sigma_{xp} + ad\sigma_{px} + bd\frac{\Delta x}{\Delta p}(\sigma_p)^2 \quad (3.11)$$

If the state $|\psi\rangle$ is a phase space state $|\psi\rangle = |n, X, P, \Delta p\rangle$, we have

$$\langle x \rangle = \langle n, X, P, \Delta p | x | n, X, P, \Delta p \rangle = X \quad (3.12)$$

$$\langle p \rangle = \langle n, X, P, \Delta p | p | n, X, P, \Delta p \rangle = P \quad (3.13)$$

$$(\sigma_x)^2 = \langle n, X, P, \Delta p | (x - \langle x \rangle)^2 | n, X, P, \Delta p \rangle$$
$$= (2n + 1)(\Delta x)^2 \quad (3.14)$$

$$(\sigma_p)^2 = \langle n, X, P, \Delta p | (p - \langle p \rangle)^2 | n, X, P, \Delta p \rangle$$
$$= (2n + 1)(\Delta p)^2 \quad (3.15)$$

$$\sigma_{xp} = \langle n, X, P, \Delta p | (x - X)(p - P) | n, X, P, \Delta p \rangle = i\frac{\hbar}{2} \quad (3.16)$$

$$\sigma_{px} = \langle n, X, P, \Delta p | (p - P)(x - X) | n, X, P, \Delta p \rangle = -i\frac{\hbar}{2} \quad (3.17)$$

Using relations (3.6), (3.7), (3.8), (3.9), (3.10), (3.11) and (1.2), we obtain

$$\langle y \rangle = \langle n, X, P, \Delta p | \mathbf{y} | n, X, P, \Delta p \rangle = aX + b \frac{\Delta x}{\Delta p} P \quad (3.18)$$

$$\langle k \rangle = \langle n, X, P, \Delta p | \mathbf{k} | n, X, P, \Delta p \rangle = c \frac{\Delta p}{\Delta x} X + dP \quad (3.19)$$

$$(\sigma_y)^2 = \langle n, X, P, \Delta p | (\mathbf{y} - \langle y \rangle)^2 | n, X, P, \Delta p \rangle = (2n+1)(a^2 + b^2)(\Delta x)^2 \quad (3.20)$$

$$(\sigma_k)^2 = \langle n, X, P, \Delta p | (\mathbf{k} - \langle k \rangle)^2 | n, X, P, \Delta p \rangle = (2n+1)(c^2 + d^2)(\Delta p)^2 \quad (3.21)$$

$$\sigma_{yk} = \langle n, X, P, \Delta p | (\mathbf{y} - \langle y \rangle)(\mathbf{k} - \langle k \rangle) | n, X, P, \Delta p \rangle = (2n+1)(ac + bd)\frac{\hbar}{2} + i\frac{\hbar}{2} \quad (3.22)$$

$$\sigma_{ky} = \langle n, X, P, \Delta p | (\mathbf{k} - \langle k \rangle)(\mathbf{y} - \langle y \rangle) | n, X, P, \Delta p \rangle = (2n+1)(ac + bd)\frac{\hbar}{2} - i\frac{\hbar}{2} \quad (3.23)$$

## 4. Transformation Laws of States and Wave Functions

Let $\{|x\rangle\}, \{|p\rangle\}, \{|y\rangle\}$ and $\{|k\rangle\}$ respectively the sets of the eigenstates of the operators $x, p, y$ and $k$, we have

$$\mathbf{x}|x\rangle = x|x\rangle \qquad \mathbf{p}|p\rangle = p|p\rangle \quad (4.1a)$$

$$\mathbf{y}|y\rangle = y|y\rangle \qquad \mathbf{k}|k\rangle = k|k\rangle \quad (4.1b)$$

The sets $\{|x\rangle\}$ and $\{|p\rangle\}$ are basis of the state space, then we have in particular the relations

$$|y\rangle = \int |x\rangle \langle x|y\rangle dx \quad (4.2)$$

$$|k\rangle = \int |p\rangle \langle p|k\rangle dp \quad (4.3)$$

To find the expression of the states $|y\rangle, |k\rangle$, we have to look for the expression of the components $\langle x|y\rangle$ and $\langle p|k\rangle$. Let us consider the relation (2.1), and introduce the states in this relation, we have

$$\begin{cases} \langle x|\mathbf{y}|y\rangle = a\langle x|\mathbf{x}|y\rangle + b \frac{\Delta x}{\Delta p} \langle x|\mathbf{p}|y\rangle \\ \langle x|\mathbf{k}|y\rangle = c \frac{\Delta p}{\Delta x} \langle x|\mathbf{x}|y\rangle + d\langle x|\mathbf{p}|y\rangle \end{cases} \quad (4.4)$$

$$\begin{cases} \langle p|\mathbf{y}|k\rangle = a\langle p|\mathbf{x}|k\rangle + b \frac{\Delta x}{\Delta p} \langle p|\mathbf{p}|k\rangle \\ \langle p|\mathbf{k}|k\rangle = c \frac{\Delta p}{\Delta x} \langle p|\mathbf{x}|k\rangle + d\langle p|\mathbf{p}|k\rangle \end{cases} \quad (4.5)$$

According to (2.2), $x, p$ on one hand and $y, k$ on the other hand are canonical conjugate operators. Therefore, in coordinate representation, we have

$$\langle x|\mathbf{y}|y\rangle = y\langle x|y\rangle \qquad \langle x|\mathbf{x}|y\rangle = x\langle x|y\rangle$$

$$\langle x|\mathbf{p}|y\rangle = -i\hbar \frac{\partial \langle x|y\rangle}{\partial x} \qquad \langle x|\mathbf{k}|y\rangle = i\hbar \frac{\partial \langle x|y\rangle}{\partial y}$$

and in momentum representation

$$\langle p|\mathbf{y}|k\rangle = -i\hbar \frac{\partial \langle p|k\rangle}{\partial k} \qquad \langle p|\mathbf{x}|k\rangle = i\hbar \frac{\partial \langle p|k\rangle}{\partial p}$$

$$\langle p|\mathbf{p}|k\rangle = p\langle p|k\rangle \qquad \langle p|\mathbf{k}|k\rangle = k\langle p|k\rangle$$

From these relations and (2.3), equations (4.4) and (4.5) respectively lead to the following differential equation systems

$$\begin{cases} y\langle x|y\rangle = ax\langle x|y\rangle - i\hbar b \frac{\Delta x}{\Delta p} \frac{\partial \langle x|y\rangle}{\partial x} \\ i\hbar \frac{\partial \langle x|y\rangle}{\partial y} = c \frac{\Delta p}{\Delta x} x\langle x|y\rangle - i\hbar d \frac{\partial \langle x|y\rangle}{\partial x} \end{cases} \Rightarrow \begin{cases} \frac{\partial \langle x|y\rangle}{\partial x} = -\frac{i}{\hbar b} \frac{\Delta p}{\Delta x}(ax - y)\langle x|y\rangle \\ \frac{\partial \langle x|y\rangle}{\partial y} = \frac{i}{\hbar b} \frac{\Delta p}{\Delta x}(x - dy)\langle x|y\rangle \end{cases} \quad (4.6)$$

$$\begin{cases} -i\hbar \frac{\partial \langle p|k\rangle}{\partial k} = i\hbar a \frac{\partial \langle p|k\rangle}{\partial p} + b \frac{\Delta x}{\Delta p} p\langle p|k\rangle \\ k\langle p|k\rangle = i\hbar c \frac{\Delta p}{\Delta x} \frac{\partial \langle p|k\rangle}{\partial p} + dp\langle p|k\rangle \end{cases} \Rightarrow \begin{cases} \frac{\partial \langle p|k\rangle}{\partial p} = \frac{i}{\hbar c} \frac{\Delta x}{\Delta p}(dp - k)\langle p|k\rangle \\ \frac{\partial \langle p|k\rangle}{\partial k} = -\frac{i}{\hbar c} \frac{\Delta x}{\Delta p}(p - ak)\langle p|k\rangle \end{cases} \quad (4.7)$$

The solutions of (4.6) and (4.7) give the expressions of the functions $\langle x|y\rangle$ and $\langle p|k\rangle$.

$$\langle x|y\rangle = C e^{\frac{i}{\hbar b} \frac{\Delta p}{\Delta x}(yx - \frac{ax^2 + dy^2}{2})} \quad (4.8)$$

$$\langle p|k\rangle = C' e^{-\frac{i}{\hbar c} \frac{\Delta x}{\Delta p}(pk - \frac{dp^2 + ak^2}{2})} \quad (4.9)$$

The coefficients $C$ and $C'$ in (4.8) and (4.9) are constant to be determined. Since, $x, p$ and $y, k$ are canonical conjugate operators, we have

$$\langle p|x\rangle = \langle x|p\rangle^* = \frac{1}{\sqrt{2\pi}} e^{-i\frac{px}{\hbar}} \quad (4.10)$$

$$\langle k|y\rangle = \langle y|k\rangle^* = \frac{1}{\sqrt{2\pi}} e^{-i\frac{ky}{\hbar}} \quad (4.11)$$

where * means the complex conjugate. Then, with the functions $\langle x|y\rangle$ and $\langle p|k\rangle$, we have the following relation

$$\langle p|k\rangle = \int \langle p|x\rangle \langle x|y\rangle \langle y|k\rangle \, dxdy$$

$$= \frac{1}{2\pi} \int \langle x|y\rangle e^{\frac{i}{\hbar}(ky - px)} \, dxdy \quad (4.12)$$

Using (4.8) and (4.9), and performing the integration in (4.12), we get a relation between the constants $C$ and $C'$

$$C' = -i \frac{\Delta x}{\Delta p} \sqrt{\frac{b}{c}} C \qquad (4.13)$$

Taking into account (4.8) and (4.9), relations (4.2) and (4.3) become respectively

$$|y\rangle = C \int |x\rangle e^{\frac{i}{\hbar b \Delta x} \Delta p (yx - \frac{ax^2 + dy^2}{2})} dx \qquad (4.14)$$

$$|k\rangle = C' \int |p\rangle e^{-\frac{i}{\hbar c \Delta p} \Delta x (pk - \frac{dp^2 + ak^2}{2})} dp \qquad (4.15)$$

And for the states $|x\rangle$ and $|p\rangle$, we have the normalization relation

$$\langle x'|x\rangle = \delta(x' - x)$$

$$\langle p'|p\rangle = \delta(p' - p)$$

where $\delta$ is the Dirac distribution.

$$\langle y'|y\rangle = |C|^2 \int \langle x'|x\rangle e^{-\frac{i \Delta p}{\hbar b \Delta x}(y'x' - \frac{ax'^2 + dy'^2}{2})} e^{\frac{i \Delta p}{\hbar b \Delta x}(yx - \frac{ax^2 + dy^2}{2})} dx' dx = |C|^2 \int \delta(x' - x) e^{-\frac{i \Delta p}{\hbar b \Delta x}[(y'x' - yx) - \frac{a(x'^2 - x^2) + d(y'^2 - y^2)}{2}]} dx' dx$$

$$= |C|^2 2\pi \hbar b \frac{\Delta x}{\Delta p} e^{-\frac{i}{\hbar b}(\frac{\Delta p d(y'^2 - y^2)}{\Delta x}\frac{}{2})} \delta(y' - y) \qquad (4.16)$$

$$\langle k'|k\rangle = |C'|^2 \int \langle p'|p\rangle e^{\frac{i \Delta x}{\hbar c \Delta p}(k'p' - \frac{dp'^2 + ak'^2}{2})} e^{-\frac{i \Delta x}{\hbar c \Delta p}(kp - \frac{dp^2 + ak^2}{2})} dp' dp$$

$$= |C'|^2 \int \delta(p' - p) e^{\frac{i \Delta x}{\hbar c \Delta p}[(k'p' - kp) - \frac{d(p'^2 - p^2) + a(k'^2 - k^2)}{2}]} dp' dp = |C'|^2 2\pi \hbar c \frac{\Delta p}{\Delta x} e^{\frac{i \Delta x}{\hbar c \Delta p}(\frac{a(k'^2 - k^2)}{2})} \delta(k' - k) \qquad (4.17)$$

If we choose, for the states $|y\rangle$ and $|k\rangle$, the following normalization relations

$$\langle y'|y\rangle = e^{-\frac{i \Delta p}{\hbar b \Delta x}(\frac{d(y'^2 - y^2)}{2})} \delta(y' - y) \qquad (4.18)$$

$$\langle k'|k\rangle = e^{\frac{i \Delta x}{\hbar c \Delta p}(\frac{a(k'^2 - k^2)}{2})} \delta(k' - k) \qquad (4.19)$$

We obtain for the coefficients $C$ and $C'$

$$|C|^2 = \frac{1}{2\pi \hbar b} \frac{\Delta p}{\Delta x} \qquad |C'|^2 = \frac{1}{2\pi \hbar c} \frac{\Delta x}{\Delta p}$$

To fulfill the relation (4.13), we may take

$$C = \sqrt{\frac{\Delta p}{2\pi \hbar b \Delta x}} e^{i\epsilon} \qquad C' = -i \sqrt{\frac{\Delta x}{2\pi \hbar c \Delta p}} e^{i\epsilon}$$

where $\epsilon$ is a real constant. Relations (4.8) and (4.9) become respectively

$$\langle x|y\rangle = \sqrt{\frac{\Delta p}{2\pi \hbar b \Delta x}} e^{i\epsilon} e^{\frac{i \Delta p}{\hbar b \Delta x}(yx - \frac{ax^2 + dy^2}{2})} \qquad (4.20)$$

$$\langle p|k\rangle = -i \sqrt{\frac{\Delta x}{2\pi \hbar c \Delta p}} e^{i\epsilon} e^{-\frac{i \Delta x}{\hbar c \Delta p}(pk - \frac{dp^2 + ak^2}{2})} \qquad (4.21)$$

And we have for the relations (4.14) and (4.15)

$$|y\rangle = \int |x\rangle \langle x|y\rangle dx = \sqrt{\frac{\Delta p}{2\pi \hbar b \Delta x}} e^{i\epsilon} \int |x\rangle e^{\frac{i \Delta p}{\hbar b \Delta x}(yx - \frac{ax^2 + dy^2}{2})} dx \qquad (4.22)$$

$$|k\rangle = \int |p\rangle \langle p|k\rangle dp = -i \sqrt{\frac{\Delta x}{2\pi \hbar c \Delta p}} e^{i\epsilon} \int |p\rangle e^{-\frac{i \Delta x}{\hbar c \Delta p}(pk - \frac{dp^2 + ak^2}{2})} dp \qquad (4.23)$$

Using relations $\langle y|x\rangle = \langle x|y\rangle^*$ and $\langle k|p\rangle = \langle p|k\rangle^*$, we deduce easily, from (4.20) and (4.21), the transformation laws for wave functions corresponding to a state $|\psi\rangle$

$$\langle y|\psi\rangle = \int \langle y|x\rangle \langle x|\psi\rangle dx = \sqrt{\frac{\Delta p}{2\pi \hbar b \Delta x}} e^{i\epsilon} \int \langle x|\psi\rangle e^{-\frac{i \Delta p}{\hbar b \Delta x}(yx - \frac{ax^2 + dy^2}{2})} dx \qquad (4.24)$$

$$\langle k|\psi\rangle = \int \langle k|p\rangle \langle p|\psi\rangle dp = -i \sqrt{\frac{\Delta x}{2\pi \hbar c \Delta p}} e^{i\epsilon} \int \langle p|\psi\rangle e^{\frac{i \Delta x}{\hbar c \Delta p}(pk - \frac{dp^2 + ak^2}{2})} dp \qquad (4.25)$$

Let us consider the case in which the state $|\psi\rangle$ is a phase space state $|\psi\rangle = |n, X, P, \Delta p\rangle$

$$\langle x|\psi\rangle = \langle x|n, X, P, \Delta p\rangle = \varphi_n(x, X, P, \Delta p)$$

$$\langle p|\psi\rangle = \langle p|n, X, P, \Delta p\rangle = \tilde{\varphi}_n(x, X, P, \Delta p)$$

So we have

$$\langle y|\psi\rangle = \langle y|n, X, P, \Delta p\rangle = \sqrt{\frac{\Delta p}{2\pi \hbar b \Delta x}} e^{i\epsilon} \int \varphi_n(x, X, P, \Delta p) e^{-\frac{i \Delta p}{\hbar b \Delta x}(yx - \frac{ax^2 + dy^2}{2})} dx \qquad (4.26)$$

$$\langle k|\psi\rangle = \langle k|n,X,P,\Delta p\rangle = -i\sqrt{\frac{\Delta x}{2\pi\hbar c\Delta p}}e^{i\epsilon}\int \tilde{\varphi}_n(x,X,P,\Delta p)e^{\frac{i}{\hbar c\Delta p}\left(pk-\frac{dp^2+ak^2}{2}\right)}dp \qquad (4.27)$$

By taking into account the relation (3.18), (3.19), (3.20), (3.21), and introducing the quantity

$$\begin{cases} Y = \langle n,X,P,\Delta p|y|n,X,P,\Delta p\rangle = aX + \frac{\Delta x}{\Delta p}\mathscr{b}P \\ K = \langle n,X,P,\Delta p|k|n,X,P,\Delta p\rangle = \frac{\Delta p}{\Delta x}cX + dP \end{cases} \qquad (4.28)$$

$$\Delta y = \sqrt{\langle n=0,X,P,\Delta p|(y-Y)^2|n=0,X,P,\Delta p\rangle} = \sqrt{a^2+\mathscr{b}^2}\Delta x = \hbar\frac{\sqrt{a^2+\mathscr{b}^2}}{2\Delta p} \qquad (4.29)$$

$$\Delta k = \sqrt{\langle n=0,X,P,\Delta p|(k-K)^2|n=0,X,P,\Delta p\rangle} = \sqrt{c^2+d^2}\Delta p = \hbar\frac{\sqrt{c^2+d^2}}{2\Delta x} \qquad (4.30)$$

after performing the calculations of the integral in (4.26) and (4.27) and using adequate arrangement, we obtain, the expression

$$\langle y|n,X,P,\Delta p\rangle = \sqrt{\frac{\mathscr{b}+ia}{\sqrt{a^2+\mathscr{b}^2}}}\left(\frac{a-i\mathscr{b}}{\sqrt{a^2+\mathscr{b}^2}}\right)^n e^{i\epsilon}\cdot e^{i\left[\left(\frac{(a^2+\mathscr{b}^2)d-a}{\mathscr{b}}\right)\left(\frac{y-Y}{2\Delta y}\right)^2+\frac{\mathscr{b}c}{\hbar}KY-\frac{cd(a^2+\mathscr{b}^2)Y^2}{4(\Delta y)^2}-\frac{a\mathscr{b}(c^2+d^2)K^2}{4(\Delta k)^2}\right]}\cdot\varphi_n(y,Y,K,\Delta k) \qquad (4.31)$$

$$\langle k|n,X,P,\Delta p\rangle = \sqrt{\frac{(-c+id)}{\sqrt{c^2+d^2}}}\left(\frac{d+ic}{\sqrt{c^2+d^2}}\right)^n e^{i\epsilon}\cdot e^{i\left[\left(-\frac{(c^2+d^2)a-d}{c}\right)\left(\frac{k-K}{2\Delta k}\right)^2+\frac{\mathscr{b}c}{\hbar}KY-\frac{a\mathscr{b}(c^2+d^2)Y^2}{4(\Delta y)^2}-\frac{cd(a^2+\mathscr{b}^2)K^2}{4(\Delta k)^2}\right]}\cdot\tilde{\varphi}_n(y,Y,K,\Delta k) \qquad (4.32)$$

## 5. Isodispersion Linear Canonical Transformation

From the relation (2.1) we have

$$\begin{cases} x = dy - \mathscr{b}\frac{\Delta x}{\Delta p}k \\ p = -c\frac{\Delta p}{\Delta x}y + ak \end{cases} \Rightarrow \begin{cases} X = dY - \mathscr{b}\frac{\Delta x}{\Delta p}K \\ P = -c\frac{\Delta p}{\Delta x}Y + aK \end{cases}$$

$$(x-X)^2 = d^2(y-Y)^2 + \mathscr{b}^2\left(\frac{\Delta x}{\Delta p}\right)^2(k-K)^2 - d\mathscr{b}\frac{\Delta x}{\Delta p}[(y-Y)(k-K)+(k-K)(y-Y)]$$

$$(p-P)^2 = c^2\left(\frac{\Delta p}{\Delta x}\right)^2(y-Y)^2 + a^2(k-K)^2 - ac\frac{\Delta p}{\Delta x}[(y-Y)(k-K)+(k-K)(y-Y)]$$

Then, taking into consideration (4.29) and (4.30), we get

$$\Sigma_x = \frac{1}{2}\left[(x-X)^2 + \frac{(\Delta x)^2}{(\Delta p)^2}(p-P)^2\right]$$

$$= \frac{1}{2}(c^2+d^2)\left[(y-Y)^2 + \frac{(\Delta y)^2}{(\Delta k)^2}(k-K)^2\right] - \frac{1}{2}(ac+\mathscr{b}d)\sqrt{\frac{c^2+d^2}{a^2+\mathscr{b}^2}}\frac{\Delta y}{\Delta k}[(y-Y)(k-K)+(k-K)(y-Y)]$$

$$\Sigma_p = \frac{1}{2}\left[(p-P)^2 + \frac{(\Delta p)^2}{(\Delta x)^2}(x-X)^2\right]$$

$$= \frac{1}{2}(a^2+\mathscr{b}^2)\left[\frac{(\Delta k)^2}{(\Delta y)^2}(y-Y)^2 + (k-K)^2\right] - \frac{1}{2}(ac+\mathscr{b}d)\sqrt{\frac{a^2+\mathscr{b}^2}{c^2+d^2}}\frac{\Delta k}{\Delta y}[(y-Y)(k-K)+(k-K)(y-Y)]$$

According to these relations, we remark that it is possible to have the relations

$$\Sigma_x = \frac{1}{2}\left[(x-X)^2 + \frac{(\Delta x)^2}{(\Delta p)^2}(p-P)^2\right] = \frac{1}{2}\left[(y-Y)^2 + \frac{(\Delta y)^2}{(\Delta k)^2}(k-K)^2\right] = \Sigma_y \qquad (5.1)$$

$$\Sigma_p = \frac{1}{2}\left[\frac{(\Delta p)^2}{(\Delta x)^2}(x-X)^2 + (p-P)^2\right] = \frac{1}{2}\left[\frac{(\Delta k)^2}{(\Delta y)^2}(y-Y)^2 + (k-K)^2\right] = \Sigma_k \qquad (5.2)$$

if we satisfy the relations:

$$(\Delta y)^2 = (\Delta x)^2 \Leftrightarrow a^2+\mathscr{b}^2 = 1 \qquad (5.3)$$

$$(\Delta k)^2 = (\Delta p)^2 \Leftrightarrow c^2+d^2 = 1 \qquad (5.4)$$

$$ac + \mathscr{b}d = 0 \qquad (5.5)$$

We define isodispersion linear canonical transformation as linear canonical transformation which fulfills (5.3), (5.4) and (5.5). In other words, isodispersion linear canonical transformations are of the form (2.1) and fulfilling (2.3), (5.3), (5.4) and (5.5), that is

$$\begin{cases} ad - bc = 1 \\ a^2 + b^2 = 1 \\ c^2 + d^2 = 1 \\ ac + bd = 0 \end{cases} \quad (5.6)$$

These relations (5.6) suggest the introduction of two parameters $\alpha$ and $\beta$ such that

$$a = cos(\alpha) \quad b = sin(\alpha) \quad c = sin(\beta) \quad d = cos(\beta)$$

The first relation in (5.6) becomes

$$cos(\alpha)cos(\beta) - sin(\alpha)sin(\beta) = 1 \Rightarrow cos(\alpha+\beta) = 1$$
$$\Rightarrow \beta = -\alpha + 2l\pi \ (l \in \mathbb{Z})$$

where $l$ is an integer. Then we have

$$\begin{cases} a = cos(\alpha) \\ b = sin(\alpha) \\ c = sin(\beta) = sin(-\alpha + 2l\pi) = -sin(\alpha) \\ d = cos(\beta) = cos(-\alpha + 2l\pi) = cos(\alpha) \end{cases} \quad (5.7)$$

Let us check that the last equation in (5.6) is verified

$$ac + bd = -cos(\alpha)sin(\alpha) + cos(\alpha)sin(\alpha) = 0 \quad (5.8)$$

Then, taking into account (2.1) and (5.4), we can write for an isodispersion linear canonical transformation as follows

$$\begin{cases} y = cos(\alpha)x + sin(\alpha)\frac{\Delta x}{\Delta p}p \\ k = -sin(\alpha)\frac{\Delta p}{\Delta x}x + cos(\alpha)p \end{cases} \quad (5.9a)$$

$$\Leftrightarrow \begin{cases} x = cos(\alpha)y - sin(\alpha)\frac{\Delta y}{\Delta k}k \\ p = sin(\alpha)\frac{\Delta k}{\Delta y}y + cos(\alpha)k \end{cases} \quad (5.9b)$$

and using (5.1), (5.2) and (5.6), we have for dispersion operators

$$\Sigma_y = \frac{1}{2}[(y-Y)^2 + \frac{(\Delta y)^2}{(\Delta k)^2}(k-K)^2] = \frac{1}{2}[(x-X)^2 + \frac{(\Delta y)^2}{(\Delta k)^2}(p-P)^2] = \Sigma_x \quad (5.10a)$$

$$\Sigma_k = \frac{1}{2}[(k-K)^2 + \frac{(\Delta k)^2}{(\Delta y)^2}(y-Y)^2] = \frac{1}{2}[(p-P)^2 + \frac{(\Delta p)^2}{(\Delta x)^2}(x-X)^2] = \Sigma_p \quad (5.10b)$$

The transformation laws for the coordinate and momentum operators eigenstates established in (4.22) and (4.23) become

$$|y\rangle = \int |x\rangle\langle x|y\rangle dx = \sqrt{\frac{\Delta p}{2\pi\hbar sin(\alpha)\Delta x}} e^{i\epsilon} \int |x\rangle e^{\frac{i\Delta p}{\hbar\Delta x}(\frac{yx}{sin(\alpha)} - cot(\alpha)\frac{x^2+y^2}{2})} dx \quad (5.11)$$

$$|k\rangle = \int |p\rangle\langle p|k\rangle dp = -i\sqrt{\frac{-\Delta x}{2\pi\hbar sin(\alpha)\Delta p}} e^{i\epsilon} \int |p\rangle e^{\frac{i\Delta x}{\hbar\Delta p}(\frac{pk}{sin(\alpha)} - cot(\alpha)\frac{p^2+k^2}{2})} dp \quad (5.12)$$

The transformation laws for the wave functions corresponding to a state $|\psi\rangle$ obtained from (4.24) and (4.25) become

$$\langle y|\psi\rangle = \sqrt{\frac{\Delta p}{2\pi\hbar sin(\alpha)\Delta x}} e^{i\epsilon} \int \langle x|\psi\rangle e^{\frac{-i}{\hbar sin(\alpha)}\frac{\Delta p}{\Delta x}(yx - cos(\alpha)\frac{x^2+y^2}{2})} dx \quad (5.13)$$

$$\langle k|\psi\rangle = -i\sqrt{\frac{-\Delta x}{2\pi\hbar sin(\alpha)\Delta p}} e^{i\epsilon} \int \langle p|\psi\rangle e^{\frac{-i}{\hbar sin(\alpha)}\frac{\Delta x}{\Delta p}(pk - cos(\alpha)\frac{p^2+k^2}{2})} dp \quad (5.14)$$

If we choose $\epsilon$ such as

$$e^{i\epsilon} = \sqrt{\frac{b-ia}{\sqrt{a^2+b^2}}} = \sqrt{sin(\alpha) - icos(\alpha)} = e^{\frac{i}{2}(\alpha - \frac{\pi}{2})} \quad (5.15)$$

the relation (5.13) become

$$\langle y|\psi\rangle = \sqrt{\frac{1}{\hbar}\frac{\Delta p}{\Delta x}(\frac{1-icot(\alpha)}{2\pi})} \int \langle x|\psi\rangle e^{\frac{i\Delta p}{\hbar\Delta x}[cot(\alpha)(\frac{x^2+y^2}{2}) - \frac{yx}{sin(\alpha)}]} dx \quad (5.16)$$

This relation looks like a fractional Fourier transformation [13]

$$F_\alpha(y) = \sqrt{\frac{1-icot(\alpha)}{2\pi}} \int f(x) e^{i[cot(\alpha)(\frac{x^2+y^2}{2}) - \frac{yx}{sin(\alpha)}]} dx$$

where $F_\alpha(y)$ is the fractional Fourier transform of $f(x)$.
For the case of the wave functions corresponding to phase space states, taking into consideration the relation (5.7), the relation (4.31) and (4.32) become

$$\langle y|n,X,P,\Delta p\rangle = e^{-in\alpha - isin(\alpha)[sin(\alpha)\frac{KY}{\hbar} - cos(\alpha)(\frac{Y^2}{4(\Delta y)^2} - \frac{K^2}{4(\Delta k)^2})] + i\epsilon} \varphi_n(y,Y,K,\Delta k) \quad (5.17)$$

$$\langle k|n,X,P,\Delta p\rangle = e^{-in\alpha - isin(\alpha)[sin(\alpha)\frac{KY}{\hbar} - cos(\alpha)(\frac{Y^2}{4(\Delta y)^2} - \frac{K^2}{4(\Delta k)^2})] + i\epsilon} \tilde{\varphi}_n(y,Y,K,\Delta k) \quad (5.18)$$

We introduce the phase space state $|n,Y,K,\Delta k\rangle$ corresponding to the couple $(y,k)$ such that

$$\langle y|n,Y,K,\Delta k\rangle = \varphi_n(y,Y,K,\Delta k) \quad (5.19)$$

$$\langle k|n,Y,K,\Delta k\rangle = \tilde{\varphi}_n(y,Y,K,\Delta k) \quad (5.20)$$

Comparing (5.19) and (5.20) with (5.17) and (5.18) and taking into consideration the relation (4.28) and (5.7), we deduce the transformation law for phase space states

$|n, Y, K, \Delta k\rangle =$
$e^{in\alpha - i\sin(\alpha)[\sin(\alpha)\frac{PX}{\hbar} - \cos(\alpha)(\frac{X^2}{4(\Delta x)^2} - \frac{P^2}{4(\Delta p)^2})] + i\epsilon} |n, X, P, \Delta p\rangle$ (5.21)

Taking into account the relations (3.20), (3.21), (5.3), (5.4), (5.10) and (5.21) we easily check the following relations

$\langle n, X, P, \Delta p|(\mathbf{y} - Y)^2|n, X, P, \Delta p\rangle = (2n+1)(\Delta y)^2$ (5.22)

$\langle n, X, P, \Delta p|(\mathbf{k} - Y)^2|n, X, P, \Delta p\rangle = (2n+1)(\Delta k)^2$ (5.23)

$\begin{cases} \mathbf{\Sigma_y}|n, Y, K, \Delta k\rangle = (2n+1)(\Delta y)^2|n, Y, K, \Delta k\rangle \\ \mathbf{\Sigma_k}|n, Y, K, \Delta k\rangle = (2n+1)(\Delta k)^2|n, Y, K, \Delta k\rangle \end{cases}$ (5.24)

## 6. Conclusion

Calculations in the present work show that many interesting results can be derived through the study of linear canonical transformation in the framework of quantum mechanics. We obtain the transformation laws for mean values and dispersions, basis states, wave functions and operators. Through the transformations of basis states and wave functions using the definition of linear canonical transformation, we are led to the studies of some interesting links between operators and integral transformations. The results established related to the effect of linear canonical transformation on phase space states and dispersion operators show that there are deep and interesting links between linear canonical transformation and the phase space representation introduced in our paper [1].

In fact, according to the relations (5.10), the dispersion operators are invariant under the action of some kind of linear canonical transformation. The latters are the so called isodispersion linear canonical transformation.

We have also found that the transformation laws for wave functions in the case of isodispersion linear canonical transformation look like fractional Fourier transformation.

We conclude that the phase space representation that we have considered is a good framework for the study of linear canonical transformation.

It is expected that more interesting studies and applications will be performed from our results.